\documentstyle[prl,preprint,aps,psfig,epsfig]{revtex}
\begin{document}
\draft

\title{Coexistence or Separation of the Superconducting,
Antiferromagnetic, and Paramagnetic  Phases in Quasi
One-Dimensional  (TMTSF)$_2$PF$_6$ ?}

\author{A.\ V.\ Kornilov$^a$, V.\ M.\ Pudalov$^a$, Y.\ Kitaoka$^b$, K.\
Ishida$^b$, G.-q.\ Zheng$^b$, T.\ Mito$^b$, and J.\ S.\ Qualls$^e$, }

\address{
$^{a)}$P.\ N.\ Lebedev Physics Institute, and P.\ N.\ Lebedev Research Center in Physics,
Moscow 119991,  Russia\\
$^{b)}$Division of Material Physics, School of Engineering
Science, Osaka University,
Osaka 560-8531, Japan\\
$^{c}$ Wake Forest University, Winston-Salem, NC 27109, USA }

\maketitle

\begin{abstract}

We report on experimental studies of the phase state and the
character of phase transitions in the quasi-one-dimensional
organic compound (TMTSF)$_2$PF$_6$ in the close vicinity of the
borders between the paramagnetic metal PM, antiferromagnetic
insulator AF, and superconducting SC states. In order to drive
the system precisely through the phase  border $P_0(T_0)$, the
sample was maintained at fixed temperature $T$ and pressure
$P$, whereas the critical pressure $P_0$ was tuned by applying
the magnetic field $B$. In this approach, the magnetic field was used
(i) for smooth and precise tuning $\delta P= P-P_0$ (thanks to a
monotonic $P_0(B)$ dependence) and (ii) for  identifying the
phase composition (due to qualitatively different
magnetoresistance behavior in different phases). Experimentally,
we measured magnetoresistance $R(B)$ and its temperature
dependence $R(B,T)$ in the pressure range $(0 - 1)$\,GPa. Our
studies focus on the  features of the magnetoresistance at
 the phase transitions  between the PM and AF
phases and in the close vicinity to the superconducting transition at
$T\approx 1$K. We found pronounced history effects arising when
the  AF/PM phase border is crossed by sweeping the magnetic field:
the resistance depends on a trajectory which the system arrives at a
given point of the $P-B-T$ phase space. In the transition from the PM
to AF phase, driven by increasing magnetic field, the features of
the PM phase extends well into the AF phase. At the opposite
transition from the AF to PM phase, the features of the AF phase are
observed in the PM phase. These results evidence for a
macroscopically inhomogeneous state, which contains
macroscopic inclusions of the minority phase, spatially separated from the majority
phase.
When the system is driven away from the transition, the homogeneous state is restored;
upon a return motion to the phase boundary, no signatures of the minority phase are
observed up to the very phase boundary.
\end{abstract}

\newpage

\section{Introduction}
The interplay (co-existence, segregation, or competition) of the
magnetic spin ordering and the superconducting pairing of
electrons is in the focus of the broad research interest. These
effects are of the key importance for understanding the rich
physics of high $T_c$ superconductors, heavy fermion compounds
and also organic conductors
\cite{HTSC,heavy_fermions:AF+SC,heavy_fermions:FM+SC,chaikin96}.
Indeed, for these materials, having low-dimensional electron systems,
the phase diagrams are surprisingly similar on the plane
``pressure''$P$ - temperature $T$ (here by
``pressure'' we mean either externally applied pressure or
internal ``chemical pressure'', i.e. dopant concentration)- see Fig. 1a. The
phase diagrams for these materials include a magnetically ordered  phase,
metallic, and superconducting phases
\cite{HTSC,heavy_fermions:AF+SC,heavy_fermions:FM+SC,chaikin96}.

The origin of the superconducting phase in (TMTSF)$_2$PF$_6$
remains puzzling; there are experimental and theoretical results
pointing at a triplet mechanism of electron pairing
\cite{triplet}. Therefore, the issue of the character of the
phase border and the origin of the transitions between magnetic,
superconducting, and paramagnetic phases (caused. e.g. by
pressure changes) become of the fundamental importance
\cite{heavy_fermions:FM+SC,magnetic_field_induced_SC}.
The most general approach to the problem was
suggested by a symmetry theory, which incorporates descriptions
of the magnetic and superconducting phases by introducing a
superspin, whose three components correspond to magnetic- and two
others to superconducting- order parameters \cite{SO(5)}. In the
frameworks of this SO(5) symmetry theory, at a certain
intermediate pressure there  might exist a state in which all
components of the superspin (magnetic and superconducting) are
nonzero. Such a ``microscopically'' mixed state possesses
magnetic ordering in the superconducting state. There are indeed
experimental indications for  the existence of a local magnetic
order in the high temperature superconductors
\cite{HTSC_magnetic_ordering}. Recently, the microscopically
mixed state was experimentally found in the heavy fermion compound
CeRhIn$_5$ \cite{kitaoka2003}.

On a  different footing, the microscopically mixed state has been
suggested in Ref. \cite{yamaji82} to occur in (TMTSF)$_2$PF$_6$
due to lacking of the complete nesting over the entire Fermi
surface. On the $P-T$ phase diagram, the mixed state should occupy
a  narrow strip of pressures just below the critical $P_{SDW}$
value. At temperature $T=1.4$\,K, this region should be about 4\%
wide on the pressure scale.

One should admit an alternative possibility, where in the
vicinity of the phase border,  a macroscopically inhomogeneous
state may arise; this state incorporates inclusions of the minority
phase embedded in the majority phase. As an example, it is well
known, that the two-dimensional Mott type insulators tend to the
formation of phase-inhomogeneous states \cite{HTSC,emery_90}. It
is also known that martensitic transformations \cite{martensite}
concomitant with  phase segregated states take place in such
materials, where the free energy of electron system (including
the magnetic energy of spin ordering) depends linearly on lattice
deformation. If the inhomogeneous state with spatially separated
phases emerged on the border of the magnetic and superconducting
states (see Fig. 1a), it would have demonstrated simultaneously
magnetic and superconducting  properties, similarly to those of
the heterophase mixed state.

The mixed state, as well as the state with spatial phase-separation,
are expected to exhibit similar purely superconducting or purely
magnetic properties far away from the  phase boundary, so that their behaviors
are indistinguishable. However, in the close vicinity of the phase boundary
$(T_0,P_0)$, properties of these two types of states are different.
In principle, one may distinguish these two possibilities,
if the system is forced to cross the phase  border by  varying pressure at constant
temperature, along the vertical trajectory in Fig.~1\,a.
In particular, for the inhomogeneous state with inclusions of the minority phase,
one might expects such effects as pre-history and hysteresis:
the properties of the system at a given point of the $P - T$ phase diagram
may depend on the pathway which the system has arrived at this point, due to
a path-dependent concentration of segregated phases.
In contrast, there is no reason to expect history effects for the
mixed state.

Straightforward performing such experiment
represents a hard technical task.  Nevertheless,
such measurements of $R(T)$ at fixed pressure values have been described recently
in Ref. \cite{vuletic02}, where the $R(T)$-dependence was studied for a number of pressure
values $P_i$ in the vicinity of $P_0$. The authors of Ref. \cite{vuletic02}
observed  hysteresis effects within the AF phase
and fitted the set of the measured $R(T,P_i)$ curves using
a simple percolation model (which modeled the inclusions of one phase into another one).
It was concluded that the observed hysteresis in the $R(T)$-dependences reflects a
macroscopically inhomogeneous state, i.e. a mixture of the two phases.
We note, however, that the identification of
the AF and metallic phases is not  trivial; at zero magnetic field the  signature of the
AF state is the onset of the $R(T)$ rise with cooling. Under circumstances when the
system may contain a new phase with apriory  unknown conduction, such
procedure may be potentially ambiguous; therefore, the conclusions made in
 Ref. \cite{vuletic02} require additional verification.

In the current paper, we applied a different experimental approach.
Using the magnetic field dependence of the $T - P$ phase border for this compound,
we swept the magnetic field  at a number of fixed pressure values in the vicinity
of $(T_0, P_0)$; the magnetic field caused changes of the phase boundary and the corresponding
phase transitions between the AF and PM states. Thus, the magnetic field was
used in our experiment for both, driving the system through the phase transition (instead of pressure),
and for reliable identifying the phase content.

We observed strong prehistory effects in the resistivity (in the presence of magnetic field), similar to those
reported in Ref. \cite{vuletic02} for the $B=0$ case.
Besides, we found prehistory effects also in the character of the magnetic field dependence $R(B)$,
which occur when the system crosses the phase boundary.
These results evidence for the macroscopically inhomogeneous
heterophase state in the vicinity of the AF-PM border.
Depending on the direction of the magnetic field sweeping,
the minority phase extended across the phase border, deep into
the majority phase.  Observation of the hysteresis and
prehistory effects is not consistent with the model of the microscopically  mixed
(coexisting) antiferromagnetic and paramagnetic
states. Thereby, in the current paper
we have unambiguously determined that the transition from
the AF insulating to PM metallic phase
takes place through emergence of a macroscopically inhomogeneous  state
with spatially  separated phases. Thus, our results obtained by independent technique,
are in agreement
with the preceding data by Vuleti\'{c} et al. \cite{vuletic02}.

\section{The idea of the experiment}
For the quasi-one dimensional compound (TMTSF)$_2$PF$_6$ at zero
magnetic field, there is a narrow pressure range in the vicinity
of $P\approx0.6$\,GPa, where the two electronic phase transitions take place
as temperature decreases. Firstly, there is a transition from
the paramagnetic PM
metallic phase \cite{note_PM}  to the insulating AF phase (spin
density wave state), and further, from the AF state to the
superconducting (SC) state. Figure 1\,a shows the corresponding
phase diagram \cite{chaikin96,vuletic02,Rmax}, which incorporates the
domains of the AF, PM, and SC phases. A vertical trajectory
$P=0.54$\,GPa  on the phase diagram corresponds
to the temperature dependence of resistance $R(T)$, shown in the
inset to Fig. 1\,a. Crossings of the phase boundaries are marked
with dots on the vertical trajectory (Fig. 1\,a) and on the
measured $R(T)$ dependence.

\subsection{Traditional approach: varying $P$ and $T$}
In order to explore the character of the transition, one has to be
able to unambiguously identify the phase character and component
content in the
vicinity of the phase boundary.
Observation  of the absolute resistivity
 solely at zero field  can hardly
provide the required information, for the resistance changes
smoothly and insignificantly in the vicinity of the second order
transition. To identify the pure homogeneous PM and AF
states, one could, in principle, make use of the temperature
dependence of conduction, which has an activated character in
the AF phase and diffusive character with the ``metallic'' sign
$d\sigma/dT <0$ in the PM phase. In practice, however, this would
require $R(T)$ measurements over a broad temperature range, which
is inaccessible in the AF phase. Indeed, for the most interesting regime
in the vicinity of the contiguity of the three phases, $T_0 =
1.3$\,K and $P_0 = 0.61$\,GPa, the temperature range of the
existence of the AF phase is limited both, from the high and low-side
(see Fig. 1\,a). Besides, direct studies of such transition by
changing the pressure {\it in situ} (i.e. along horizontal trajectories in Fig. 1\,a) at low
temperatures represent a very hard
technical task.

\subsection{Alternative approach: varying $P_0$ and $T_0$ }

According to our measurements in magnetic field and the earlier results
(see, e.g., Ref. \cite{P0(B)}, the AF/PM border $T(P)$ shifts to higher
temperatures as magnetic field grows. Figure 1\,b shows
schematically the changes of the border  with magnetic field. Due
to the smooth and monotonic dependence of $T_0$ on  magnetic
field, this dependence may be used for varying $T_0$.  Thus,  the system
may be forced  to cross the border by varying the magnetic
field at fixed values of pressure and temperature. Figure 1\,b
shows that when the initial $P$ and $T$ values (at zero field) are
chosen in the vicinity but bigger than $P_0$, $T_0$, the phase
trajectory of the system (trajectory 2) will cross the phase
border with magnetic field growth. Thus, crossing the border
occurs due to the changes in $T_0(B)$ and $P_0(B)$ at fixed $T$ and
$P$ values. Besides, crossing the border in the presence of magnetic field
causes qualitative changes in the  behavior of magnetoresistance,
which are used in the current work for identifying  the phase
state and phase content of the system.

\section{Experimental}
Two samples - (TMTSF)$_2$PF$_6$ single crystals have been grown
using a conventional electrochemical technique; the typical sample
sizes were $2\times 0.8\times 0.3$\,mm$^3$, along the crystal
directions {\bf \em a}, {\bf \em b}, and {\bf \em c},
correspondingly. The two nominally equivalent samples
showed qualitatively similar behavior and had slightly different
resistivity value at low temperature.
Measurements were made by four probe ac lock-in
technique at 32\,Hz frequency. For electrical contacts, four
$25\mu$m Pt-wires were attached by a graphite conductive paint to
the sample  along the most conducting direction {\bf \em a} at the
{\bf \em a-c} plane. The sample and a manganin pressure gauge
were  placed inside a nonmagnetic  pressure cell
\cite{cylinder_cell} filled with Si-organic (polyethilenesiloxane)
PES-1 pressure transmitting liquid; a required pressure was
created at room temperature. The Ohmic character of the contacts
to the sample was confirmed by the negligibly small out-of phase
component of the measured voltage drop between the contacts.

The pressure cell was mounted in a cryostat
in the bore of a 16\,T superconducting magnet.  Measurements at
temperatures $T\geq 1.4$\,K were done in the cryostat with $^4$He
pumping.
For all measurements, the
magnetic field was applied along the least conducting direction,
{\bf \em c}, of the crystal and current was applied along {\bf \em
a}. Temperature of the pressure cell was measured using the
RuO$_2$ resistance thermometer, and heat contact of the sample to the
liquid helium (or mixture) bath was provided with Pt-wires.
In order to implement the idea of measurements with crossing the phase border
 for the account of magnetic field changes, the
pressure value has to be  set in the interval
$0.62-0.64$\,GPa at $T\approx 1.4$\,K.

\section{The results of measurements}
The magnetic field dependence of the resistance is qualitatively
different for three different trajectories ({\it 1, 2, 3}) in the
$B-P-T$ phase space, depicted on Fig.~1\,b.

\subsection{Trajectory {\em 1}}
Figure 2\,a shows a typical example of the magnetic field dependence of the
resistivity, corresponding to the trajectory {\it 1} on Fig. 1\,b.
For such trajectory, which entirely belongs to the AF domain,
changes in the resistivity are not accompanied with hysteresis.
Resistance increases monotonically with magnetic field; in strong field,
the so called ``Rapid oscillations'' (RO) appear on the background of the monotonic
$R(B)$ growth \cite{note_RO}; such $R(B)$ dependence is typical for the AF phase
\cite{uji97}. The oscillations can be more clearly seen in the  derivative
$dR(B)/dB$, shown in the inset to Fig.~2\,a. As pressure increases
(but still remains less than the critical $P_0(B)$ value), the resistance magnitude
decreases, whereas $R(B)$ dependence does not change qualitatively.

\subsection{Trajectory {\em 3}}
When the initial $T,P$ values are chosen essentially larger than $T_0, P_0$,
the trajectory {\em 3} of the system (Fig.~1\,b) lies  entirely in the PM domain
over all range of the magnetic field changes.  Figure 2\,b shows that the
magnetoresistance in this case has a character qualitatively   different
from that discussed above for the AF phase. As magnetic field increases,
the smooth growth of $R(B)$ transforms into step-like changes,
which are related to the developing cascade of
transitions between the states with different nesting vector
\cite{chaikin96,cooper89,hannahs89,kornilov02}.
In strong fields and at low temperatures, the  transitions between the states
with different nesting vector have
a  character of the 1st order phase  transitions \cite{kornilov02,lebed}.
Correspondingly, the step-like changes of $R(B)$ in strong fields are accompanied
with hysteresis in $R(B)$ \cite{kornilov02}. Such hysteresis may be noticed in
$R(B)$ traces in strong fields, presented in Fig.~2\,b.

The inset to Fig. 2\,b shows on the $B-T$ plane the corresponding phase diagram,
which includes different sub-phases of the field-induced spin density waves
\cite{chaikin96,cooper89,hannahs89,kornilov02}. On Figure 2\,b, the $R(B)$
curve at $T=Const$ corresponds to the isothermal  trajectory on the $B-T$
phase diagram (shown in the inset to this figure) which sequentially crosses
different sub-phases. The corresponding jumps in $R(B)$ are periodic in $1/B$
\cite{hannahs89,cooper89} and well correspond to the phase boundaries on the
known $B-T$ phase diagram of the FISDW-regime (see inset to Fig. 2\,b)
\cite{kornilov02}.

Each individual subphase has its own nesting vector and the quantized Hall
resistance value \cite{cooper89,hannahs89,yakovenko}.
Indices $N$ for different subphases in Fig. 2\,b correspond to the number of
filled Landau levels in the quantum Hall effect, and, simultaneously, determine
quantized changes of the nesting vector \cite{lebed}. When pressure $P$ decreases
(but still remans bigger than the critical $P_0$ value), the resistance magnitude
smoothly increases whereas $R(B)$ dependence does not change qualitatively.
The resistance jumps related to the phase transitions persist and
monotonically shift to lower fields, thus indicating the shift of the phase boundaries
\cite{kornilov02}.

\subsection{Trajectory {\em 2}}
When the initial $P,T$ values are chosen in the vicinity but slightly bigger than
$P_0, T_0$, one can expect the phase trajectory {\em 2} to cross the border with
increasing magnetic field, as discussed above. The inset to Fig.~3\,a shows the
$R(T)$ dependence measured at $B=0$; it evidences for the true metallic initial
state of the sample at $P=0.64$\,GPa (which is close to the critical value
$P_0\approx 0.61$\,GPa). When magnetic field increases, the resistance changes
insignificantly up to $B \approx 7$\,T (see Fig.\ 3a). Upon further increase of $B$
up to 16\,T, the resistance sharply grows by 3 orders of magnitude. This growth
indicates the transition from the metallic PM to insulating AF phase. On
the background of the growing monotonic resistivity component, one can
note the appearance of non-monotonic periodic variations
of resistance (starting from $B\approx 8$\,T), which are absolutely
untypical for the AF phase.

As  magnetic field is swept down (i.e. decreases from 16\,T to 7\,T),
a strong hysteresis ($\sim 20\%$) is revealed in the resistance (Fig.~3\,a),
whereas the  non-monotonic component of $R$ practically disappears. The $R(B)$
hysteresis and the appearance and disappearance of the non-monotonic component
of resistance depend only on the absolute magnetic field value
and do not depend on its direction (compare $R(B)$ and $R(-B)$ in Fig.~3\,a). The
magnitude of hysteresis grows with magnetic field. Upon repeated magnetic field
sweeps from 0 to 16\,T, the described above $R(B)$ dependence reproduces fully.
The non-monotonic component of the resistance is more clearly seen in the
$dR/dB$ derivative shown in Fig.~3\,b. It is worthy of note that the non-monotonic
component is observed only when magnetic field is increased and is practically
invisible when magnetic field is decreased from 16\,T.

Vertical arrows in Fig.~3\,b depict the locations of
the FISDW phase boundaries, which  were experimentally determined in Ref.
\cite{kornilov02} from the jumps in $R(B)$ versus  field in the FISDW area of the pure PM phase.  The
location of peaks in $dR/dB$  in Fig.\ 3\,b coincides well with the arrows
(i.e. with the anticipated borders of the FISDW phases). The inset to Fig.~3\,b
demonstrates that these peaks are equidistant in $1/B$. For these two reasons,
we may identify the observed peaks in $dR/dB$ with crossing the borders between the
FISDW phases with $N= 6 \Longleftrightarrow 5$, $5\Longleftrightarrow 4$,
$4 \Longleftrightarrow 3$, and $3 \Longleftrightarrow 2$, correspondingly,
on the phase diagram Fig.~2\,b upon isothermal sweeping the field.

We wish to stress once more that the existence of the peaks and their periodicity in
$1/B$ would be quite natural for the PM state, but is absolutely unexpected for
the AF state. On the $dR/dB$ dependence, the next
peak ($N=2 \Longleftrightarrow 1$)
at $B\sim 14$\,T is not seen, despite the peak amplitude is known to strengthen  with
decreasing $N$ \cite{kornilov02}. The absence of this peak indicates almost complete disappearance
of the PM phase and the onset of the homogeneous AF state at the field of $B=14$\,T.
As an additional confirmation of this conclusion, we note that in stronger fields $B>14$\,T rapid oscillations
may be seen in Fig.~3\,b; these RO oscillations are characteristic for the
AF phase \cite{note_RO}.

Magnetic field dependences of the derivative $dR/dB$ are shown in
Fig.~4 for different temperatures. One can see that
the hysteresis of $R(B)$  in fields sweeping up and down disappears as
temperature increases. Note that at low temperature, the
hysteresis reveals itself not only in the magnitude of $R(B)$
(and $dR/dB$) but also in the qualitatively different character
of the $R(B)$ dependence. When the field is swept up,
the $R(B)$ dependence exhibits jumps (marked by arrows in fields $B=8-12$\,T). The
jumps are characteristic for the FISDW phase transitions in the PM
phase, whereas the system passed transformation to the AF phase
starting from field $B\approx 6$\,T; the strong growth of
resistance in Fig.\ 3\,a and  the appearance of RO evidence for
this transformation. When the magnetic field is  swept down,
these anomalous jumps are almost invisible and one can see only the anticipated
RO oscillations \cite{note_RO}.

\section{Discussion of the results}

The most essential results of our studies are as follows:\\
1) As expected, when the field increases (decreases) along the trajectory {\em 2}
(see Fig. 1\,b), the system exhibits the phase transition. The steep, a factor of $10^3$, raise (fall)
of the resistance at $B\approx 6$\,T evidence for this phase transformation.\\
2) At the transition from the PM to AF phase, rather far away from the phase border, the
magnetoresistance continues  to exhibit residual signatures of the metallic (minority)
phase. When the field grows, the signatures of the minority phase almost disappear
and are not restored when the system approaches back to the same phase border (i.e. as field decreases).
In other words, at such transition, a strong hysteresis is observed both, in the magnitude
of $R$ and in the character of $R(B)$ dependence.\\
3) Upon return transformation from the AF to PM phase (with decreasing field), a hysteresis
in the magnitude of $R$ is observed:  at the transition, the resistance is noticeably higher than
that for the pure metallic PM phase (or than the resistance value measured as the field grows from $B=0$).
The ``true''
value of $R$ is restored only when the field is decreased to zero.

In view of the complicated character of the magnetoresistance
behavior, which exhibits signatures of both phases, the
experimental determination of the AF and PM phases becomes of the
principle importance. According to the existing theory
\cite{brazovski}, the SDW-PM transition is expected to be either of
the second, or weak first order. Experimental data are in
agreement with this conclusion \cite{specific_heat}.
In the vicinity of the critical pressure the SDW gap $\Delta$, in
general, might be small as compared to the antinesting parameter
$t_b^\prime$ \cite{brazovski}. In this case, the pseudo-activated $R(T)$ dependence
in the SDW phase would have a semi-metallic character and would be
indistinguishable from the ``metallic'' $R(T)$ dependence, thus
making the identification of the two phases difficult. However,
 for the specific 2D tight binding case in (TMTSF)$_2$PF$_6$,  $\Delta$
does not depend on pressure in the vicinity of the critical
pressure and is not small at the transition \cite{brazovski}. This
agrees with experimental observations \cite{danner1996}, where
$R(T)$ was shown to have a pseudo-activated character with rather
big gap in the vicinity of the critical pressure. We use
therefore, the sharp growth of $R(T)$ (starting from $B\approx
7$T in Fig.~3\,a) as a firm indication of the onset of the SDW
phase.

\subsection{Inhomogeneous state: phase separation or phase mixing ?}

Manifestly, our experimental results do not fit the behavior, anticipated for a microscopically
mixed state made of the two coexisting phases. For such a state, the hysteresis effects and the
dependence of the phase content on the prehistory, should not occur. The behavior
described above is also not typical for a homogeneously ``overheated'' or ``overcooled''
phases at the first order phase transitions, for the minority phase disappears smoothly
with no sharp jumps in $R$. Besides, for the 2nd order transition
 in a homogeneous system, neither hysteresis nor ``overheating/overcooling''
should take place. In the domain of the phase space, where only PM or AF phase should  exist,
clear  signatures of the opposite phase are observed  beside the ``correct'' phase. Therefore,
the appearance of the hysteresis  and the distinct signatures
of the presence of both phases in the same domain of the phase space, both evidence
that the  phase content of the system becomes inhomogeneous. From a theoretical viewpoint,
the cascade of transitions could also exist in the AF phase (accompanied with a corresponding
jumps in $R(B)$ \cite{brazovski}. However, such cascade has never been observed experimentally
in the AF phase. Furthermore, even if the cascade of transitions occurs,
as a homogeneous state, it would not  give
raise to pre-history effects such as observed in our experiment.

The phase-inhomogeneous state is not a consequence of the  inhomogeneity
of the sample or of the external pressure. The experimental results which prove this are as follows:\\
1. The phase-inhomogeneous state was observed on two different samples; the hysteresis and
prehistory effects were qualitatively similar in both samples
(compare Figs.~3\,b  and 4).\\
2. The existence of the prehistory in the  appearance of the phase-inhomogeneous state
contradicts the assumption of the inhomogeneity of the sample or  external pressure.
Indeed, if such inhomogeneities exist, they would manifest always,
and would not disappear at the field sweeping through the border of the
2nd order transition; therefore, the prehistory effects would not  take place.\\
3. Hysteresis in the character of $R(B)$ dependence arises only at pressure and
temperature values in the vicinity of the phase border ($T_0,P_0$).
No history effects are observed  when the system is moved away from the phase
boundary in either pressure or temperature axes. This may be seen, e.g. in Fig.~4,
where in strong fields  $B>12$\,T, rapid oscillations \cite{note_RO}
have the same magnitude and phase for  the field sweeping up and down. \\
4. Low residual value  of the resistivity,
$60\times 10^{-6}$\,Ohm$\cdot$cm (see Fig.~3\,a), evidences for high quality
of  the samples.

One could assume that  the phase-inhomogeneous state arises
due to a positive  surface energy at the  border between AF and PM phases,
and therefore, the two phases are spatially separated. On the contrary,
the microscopic coexistence of the two phases would require a negative surface
energy. However, the existence of a noticeable surface energy would  mean that
this transition is of the well-pronounced 1st order; such assumption seems  to
disagree with theoretical predictions for the SDW
phase transition \cite{brazovski} and experimental data \cite{specific_heat}.

\subsection{Prehistory effects}
The prehistory effect is the most unexpected among the results
obtained, even more unexpected than the hysteresis.
This phenomenon is illustrated on
Fig.~5\,b, where four different dependences $dR(B)/dB$ are
compared; these dependences correspond to 4 trajectories ($AB,
BC, CD$, and $DE$) shown in Fig.~5\,a. When the system crosses
the PM-AF phase border (at $B\approx 6$\,T) and moves deep into
the AF domain along the trajectory $AB$, the derivative
$dR(B)/dB$ exhibits peaks (marked  with vertical arrows). These
peaks correspond well to the resistance jumps observed in the PM
phase at crossing the borders between the FISDW sub-phases
$4\Longleftrightarrow 3$, and $3\Longleftrightarrow 2$
\cite{kornilov02}; correspondingly, the peaks have nothing in
common with the AF phase in which the system is supposed to be
for the  given  $P, B, T$ values. The existence of these peaks
 evidence that, at least, a part of the sample did not
 transform into the insulating
 AF phase and remains in the metallic PM phase.
 In the field $B\approx 15$\,T, instead of the next
 anticipated peak (which would correspond to the FISDW
 transition $2\Longleftrightarrow 1$),
 one can see only weak oscillations reminiscent of the
 RO oscillations in the AF phase.
This points out at almost complete disappearance  of the PM  phase
and the onset of the  homogeneous  AF state.

\subsection{Phase separation at zero magnetic field}
In the described above experiments, the presence of the magnetic field was not of a
principle importance. The role of the magnetic field was to produce a qualitative difference
between the $R(B)$ dependences in the AF and PM phases; this is necessary to identify crossing
the border and to reveal the phase content of the inhomogeneous state. In our view,
the phase-inhomogeneous state with inclusions of minority phase imbedded into the
majority phase must also arise in the transition from PM to AF phase with decreasing temperature
(see the phase diagram in Fig.~1a). In this case, however, the  resistance changes are anticipated
to be weak and of a quantitative rather than qualitative character. Such measurements have been already
undertaken in Ref. \cite{vuletic02}, and our task was to test or confirm these results for same
samples in which we have determined the character of the transition in non-zero magnetic field.

For the experiment we have chosen the pressure $P=0.5$\,GPa, which is less than the
critical value. The results are represented in Fig.~6. For this pressure and $B=0$,
 the system transforms from the metallic to AF state as temperature decreases
below $T = 7$\,K \cite{chaikin96,vuletic02}. At the transition, the resistance sharply raises
 and further grows with decreasing the temperature; this behavior corresponds to
the onset of the insulating state (spin density wave). The variations of the resistance with
 temperature along the trajectory $AB$ are shown in Fig.~6\,a. The final resistance value
 at point $B$ corresponds to the minimal temperature 4.2\,K in this experiment.

 According to the above assumption, at point $B$ the system
 has a spatially inhomogeneous phase
 content: beyond the majority insulating AF phase, it contains
 also inclusions of the minority metallic
PM phase. The following experiment was done in order to check
this assumption: the magnetic field was increased  from 0 to
16\,T; according to the above results, such strong field should destroy
completely the inclusions of the minority phase. Figure~6\,b shows
resistance changes with increase (trajectory {\it BC}) and
subsequent decrease (trajectory {\it CD}) of the magnetic field.
After magnetic field is decreased to zero, the system returns to
a state  (point $D$) similar to the initial (point $B$). However,
magnifying the data in the inset to Fig.~6\,b reveals a small
($\sim 5\%$) increase in the resistance at point $D$ as compared
to that at the initial point $B$. This minor difference evidences
for decreasing the share of the well-conducting metallic phase.
The observed  hysteresis is weak, therefore its interpretation
could hardly  be done without preliminary studies of much
stronger hysteresis effects in magnetic field. The hysteresis in
our $R(T,B=0)$ measurements is essentially weaker than that in Ref.
\cite{vuletic02}. The reasons for this might be related to a
somewhat smaller deviation of  the pressure from the critical $P_0$
value in Ref.~\cite{vuletic02} than in our studies. Indeed, the
lower transition temperature $T_{\rm SDW}=2.5$\,K  in Ref.
\cite{vuletic02} as compare to $T_{\rm SDW} = 7$\,K in our
studies   (see Fig.~6\,a) indirectly indicates   for such
difference.

As the field varies repeatedly from $D$ to $C$ and back, no
irreversible changes in the resistance are observed; we conclude the
inclusions of the metallic phase disappeared. Upon
further increase of temperature at zero field, the resistance
varies along the trajectory $DA$. Repeated coolings reproduce the
trajectory $AB$ within 1\% accuracy, the result which evidences
for restoring the phase-inhomogeneous state.

\section{Conclusion}
The experiments described above reveal
hysteresis in the magnitude of the resistance and the character of its
variation with magnetic field, which develop at the transition from metallic to
antiferromagnetic insulator state. Furthermore, we found that the behavior of
the resistance with magnetic field becomes prehistory dependent.
These results evidence unambiguously for the occurrence
of the inhomogeneous state in the vicinity of the phase boundary between
the PM and AF phases; we conclude that this state consists of inclusions of the
minority phase imbedded into the majority phase. No data are available yet
on the spatial arrangement of the two-phase state. The conclusions we drawn,
do not depend on any model assumptions about the spatial arrangement of the two-phase state,
because for identifying the phase content we used qualitative difference
in the magnetoresistance behavior in AF and PM states.
Our results are in a good agreement with previous data \cite{vuletic02},
 obtained in a different way.

The observed phenomena of the phase separation at the second
order transitions are similar to those typically seen in
martensitic transformations. If this analogy is not accidental,
it suggests that  the free energy of the electron system depends
linearly on the charge or SDW distortion. This conclusion requires
experimental verification.

The minority phase inclusions can be
eliminated
completely as the system
moves far away from the boundary, deep in the majority phase domain. The hysteresis
in the magnitude and in the field dependence of the resistance does not depend on time;
it is a stationary and well reproducible effect. The most striking  evidence for
 the heterophase content was obtained from experiments in finite magnetic fields.
 However, similar
phase-inhomogeneous state has been also observed in zero field,
for the transition from metallic to
antiferromagnetic insulator (SDW)-state. These observations confirm that
the magnetic field does not play an essential role in the occurrence of the
heterophase state. Extending this analogy to the
superconducting transition, we note
an interesting possibility that  the transition from the antiferromagnetic
insulator to superconductor state might also occur via superconducting transition
in inclusions of the minority metallic phase, rather than between the two homogeneous
AF and SC states. This suggestion also requires  an additional experimental verification.

\section {Acknowledgements} The authors are grateful to
V.\ Kiryukhin, V.\ Podzorov, S.\ Brazovski, A.\ G.\ Lebed, and
V.\ Yakovenko for discussions.  The work was partially supported
by INTAS, RFBR, NWO, Russian Federal Programs ``Quantum
macrophysics'', ``Strongly correlated electrons'',
''High temperature superconductivity'', ``Integration of high education and academic
research'', ``The State support of the leading scientific
schools'', and COE Research in Grant-in-Aid for Scientific
Research, Japan.

\newpage

{\Large Figure captions}\\

Fig.~1. a) The $P-T$ phase diagram for (TMTSF)$_2$PF$_6$ in the
absence of magnetic field. Difference phases are designated as
follows: PM - paramagnetic metal, AF - antiferromagnetic
insulator, SC - superconductor. Vertical line shows the isobaric
trajectory $P=0.54$\,GPa, which crosses the PM - AF and AF - SC
phase borders. The inset shows the temperature dependence of the
resistance at $P=0.54$\,GPa, which corresponds to the vertical
trajectory at the main panel. Two bold dots at the $R(T)$ curve
and at the trajectory mark two corresponding phase transitions,
from the metallic PM to the insulating AF state, and from AF to
the superconducting SC state. b) Schematic $P-B-T$ phase diagram
for the nonsuperconducting phase space  ($T \geq 1.14$\,K).

Fig.~2. a) Variation of the resistance with magnetic field in the AF state, which corresponds
to the motion along the trajectory {\em 1} in Fig.~1\,b. In the inset, on the derivative
$dR/dB$, one can see the RO oscillations, which are typical for the  AF phase.
b) Variation of the resistance with magnetic field in the PM state, (trajectory {\em 3}
in Fig.~1\,b). In high fields, one can see jumps in $R(B)$, which are typical for the
PM state and correspond to crossing the boundaries between the FISDW phases with different
nesting vector. The inset shows the corresponding phase diagram for the FISDW region,
experimentally determined in Ref. \protect\cite{kornilov02}.

Fig.~3. a) Variation of the resistance with magnetic field for the case of crossing the
PM-AF boundary (trajectory {\it 2} in Fig.\ 1\,b). The inset shows $R(T)$ changes
in the initial  state at $B=0$, characteristic for the PM metal ($dR/dT>0$).
The sharp growth of $R(B)$ at $B\approx 7$\,T corresponds to the PM $\rightarrow$ AF transition.
b) Magnetic field dependence of the derivative $dR/dB$, which corresponds to the $R(B)$ curves in Fig.~3\,a
upon increasing and decreasing field. The vertical arrows show the borders between the FISDW phases, depicted
from the experimentally determined phase diagram (see \protect\cite{kornilov02} and the inset to Fig.~2\,b).
The inset demonstrates the periodicity of the $dR/dB$ peaks in $1/B$. Sample \#1.

Fig.~4. Magnetic field dependence of the derivative $dR/dB$ for four temperatures. The curves show
rapidly disappearing  hysteresis  as temperature grows, and a characteristic nonmonotonic temperature
dependence of the rapid oscillations. Sample \#2.

Fig.~5. History effects developing in $R(B)$ (panel a) and in $dR/dB$ (panel b)   as magnetic field varies
along the trajectories {\it AB, BC, CD}, and {\it DE}. Two arrows mark the peaks in $dR/dB$
corresponding to the FISDW transitions in the PM phase; these peaks are missing as field decreases.

Fig.~6. a) History effects in the temperature dependence  $R(T)$ at zero field
when the trajectory crosses the PM-AF phase boundary; b) Re-establishing
the homogeneous state by sweeping the  magnetic field to 16\,T and back to zero.
Pressure is $P=0.5$\,GPa.

\begin{figure}
\centerline{\psfig{figure=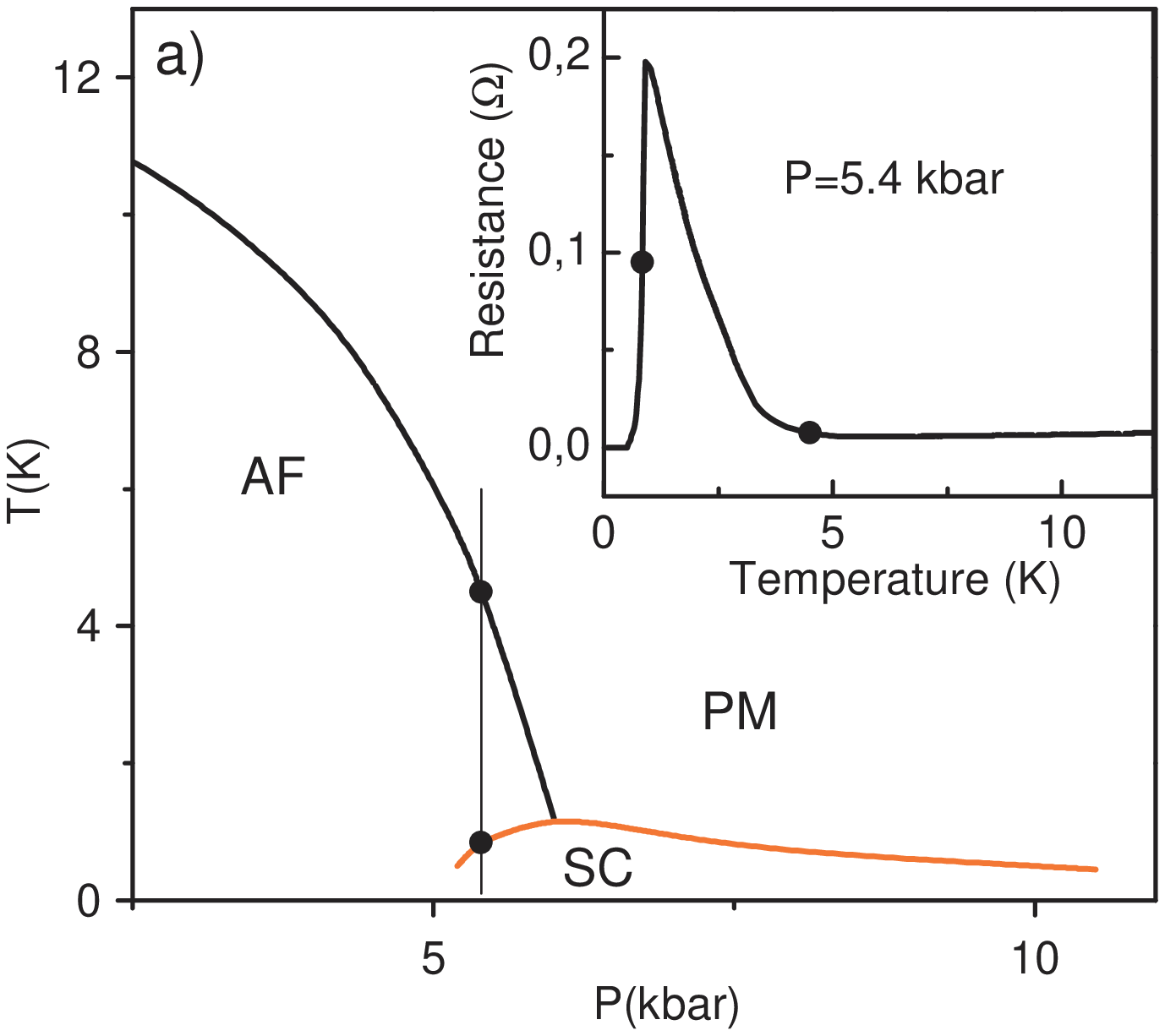,width=4.1in} } \vspace{0.7in}
\centerline{\psfig{figure=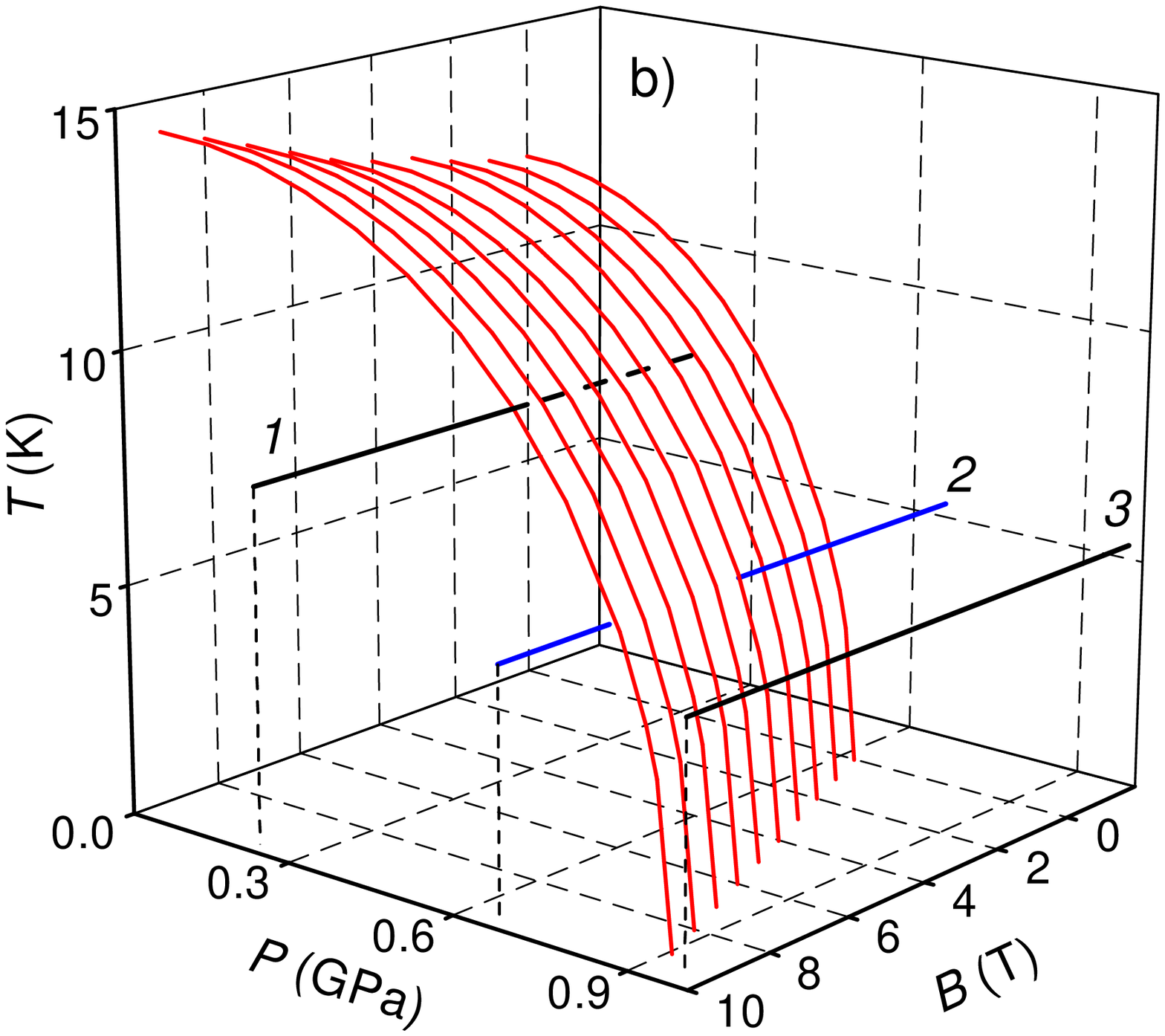,width=4.1in} } \vspace{0.3in}
\caption{}
\label{fig1}
\end{figure}

\newpage

\begin{figure}
\centerline{\psfig{figure=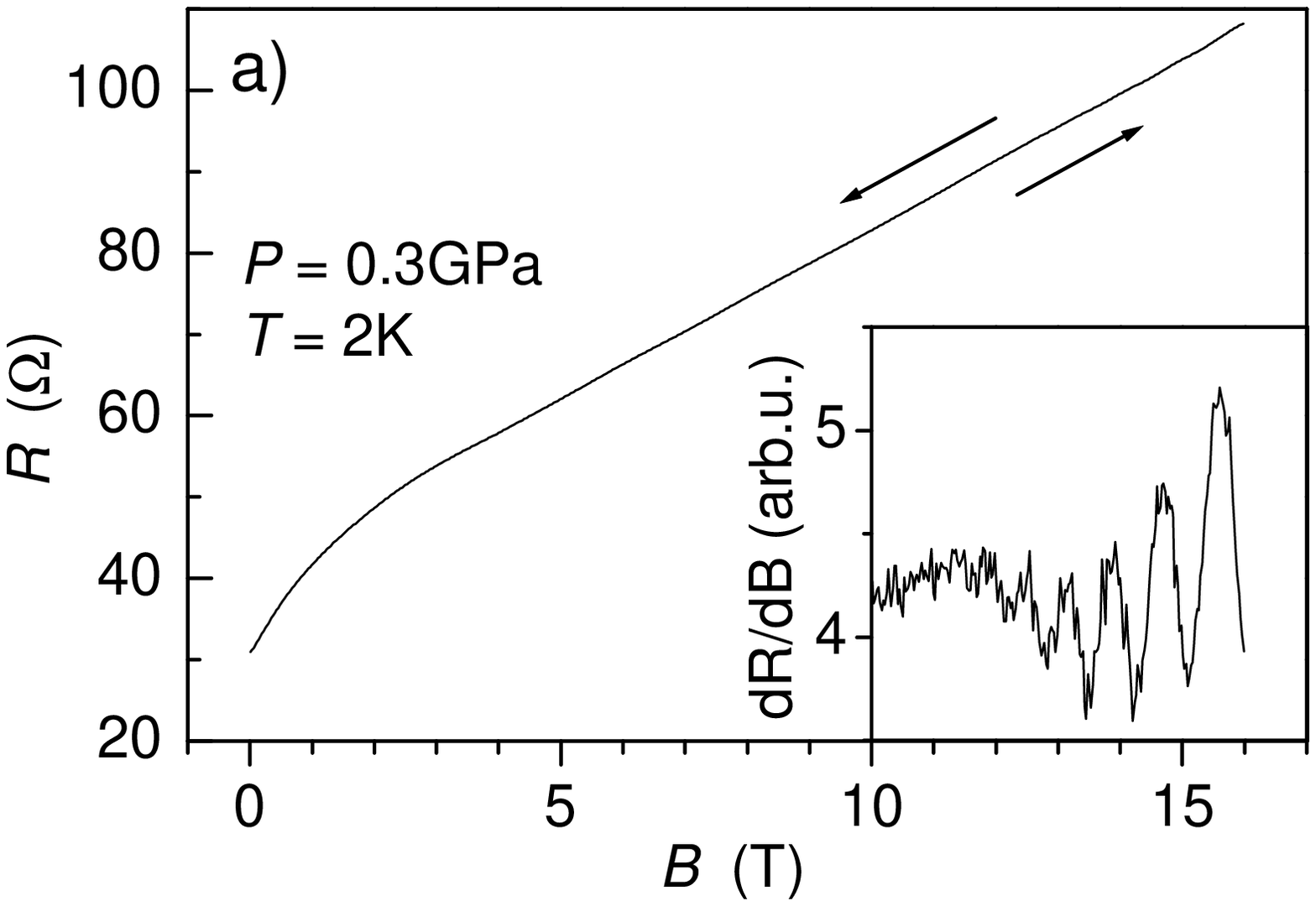,width=5.4in} }
\vspace{1in}
\centerline{\psfig{figure=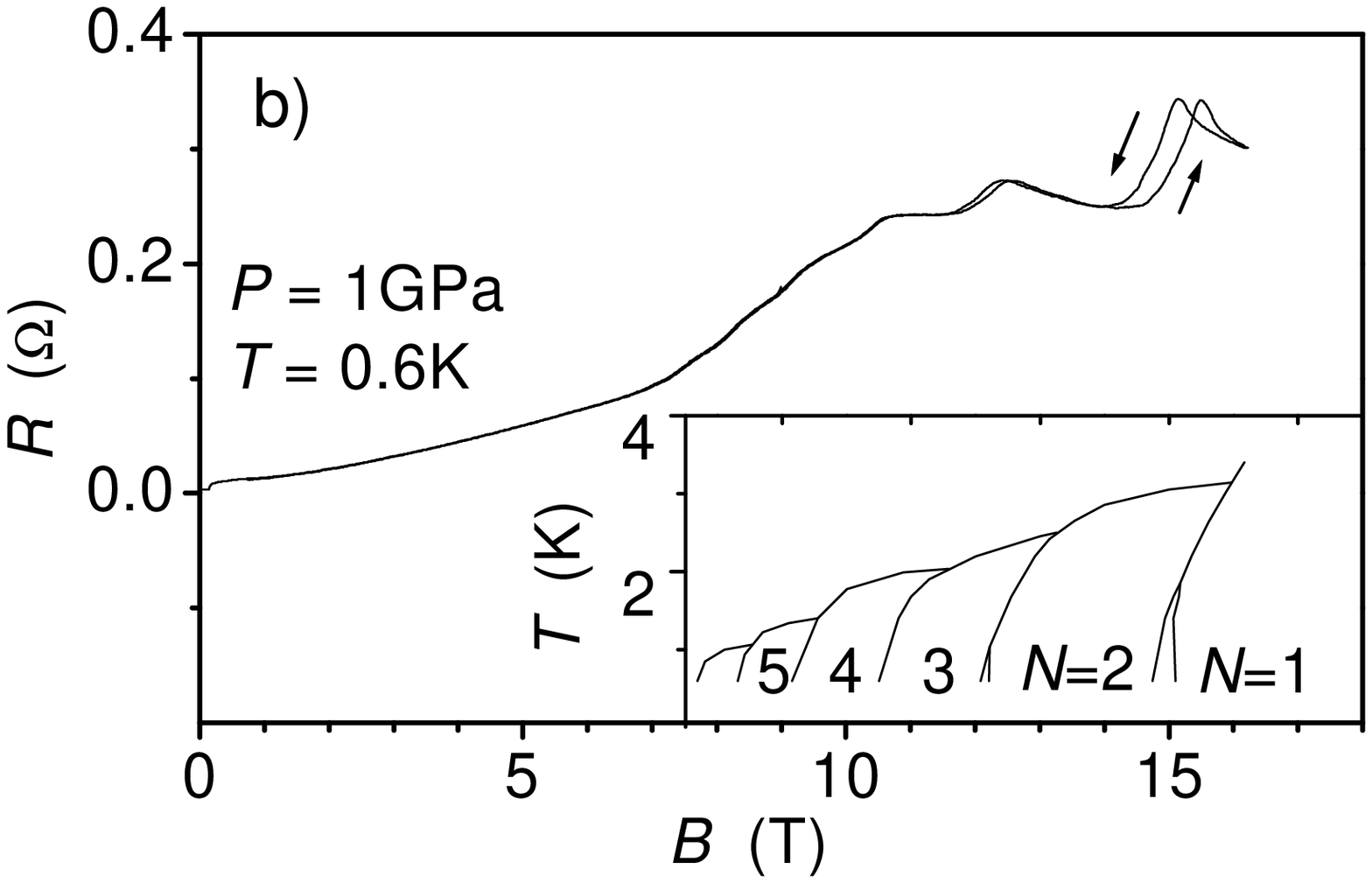,width=5.4in} }
\vspace{0.5in}
\caption{}
\label{fig2}
\end{figure}

\newpage
\begin{figure}
\centerline{\psfig{figure=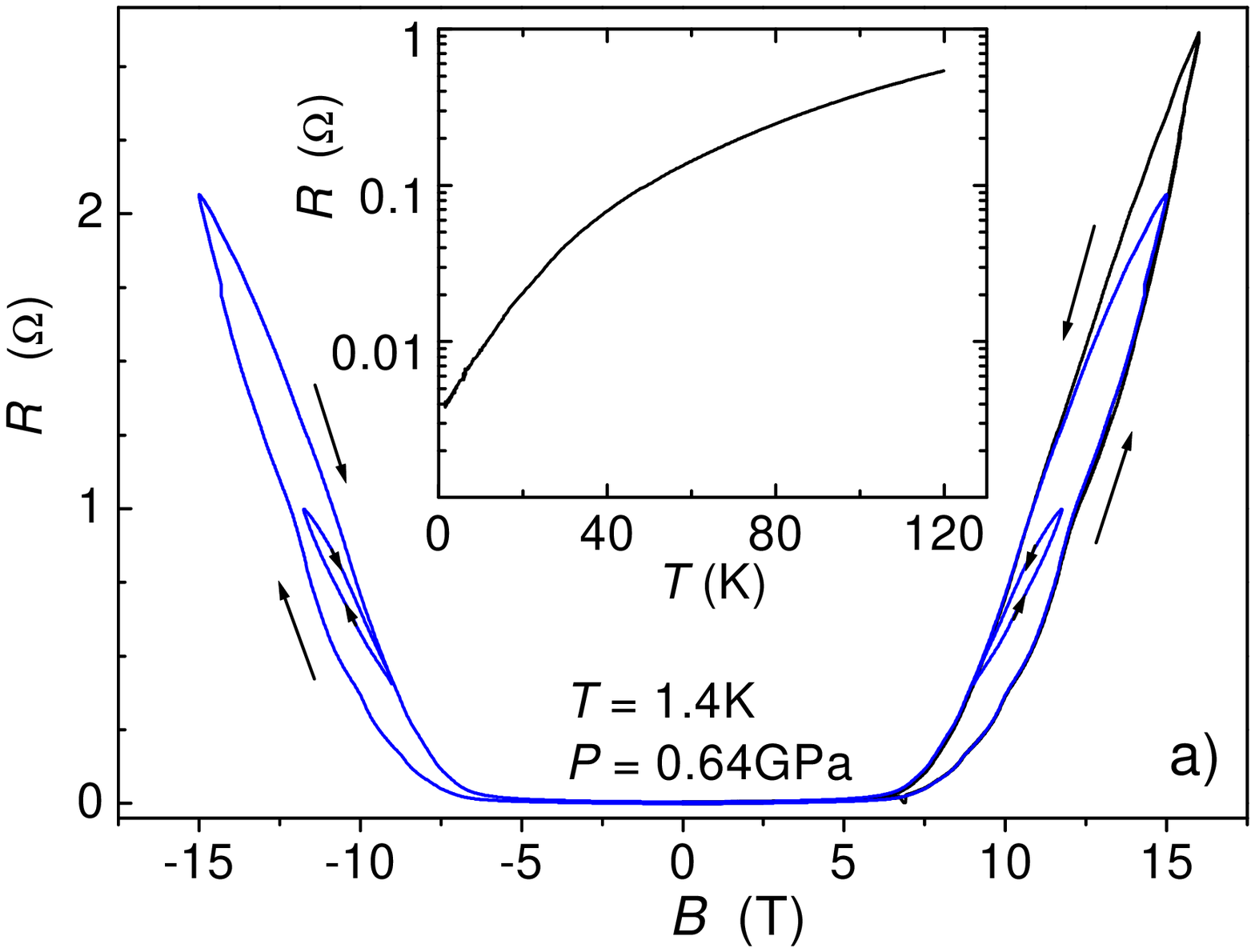,width=5.2in}}
\vspace{0.5in}
\hspace{-0.3in}\centerline{\psfig{figure=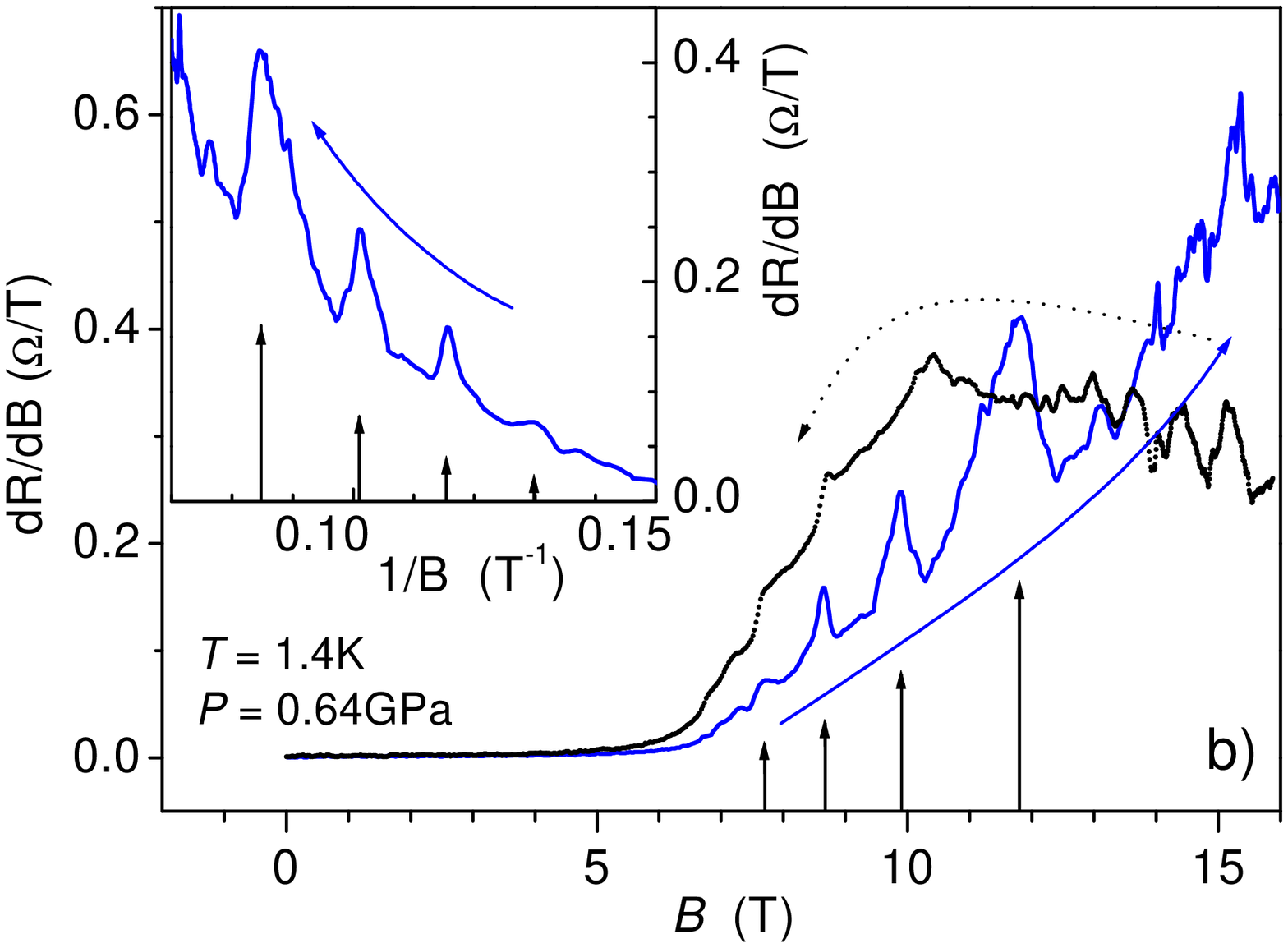,width=5.5in}}
\vspace{0.2in}
\caption{}
\label{fig3}
\end{figure}

\newpage
\begin{figure}
\centerline{\psfig{figure=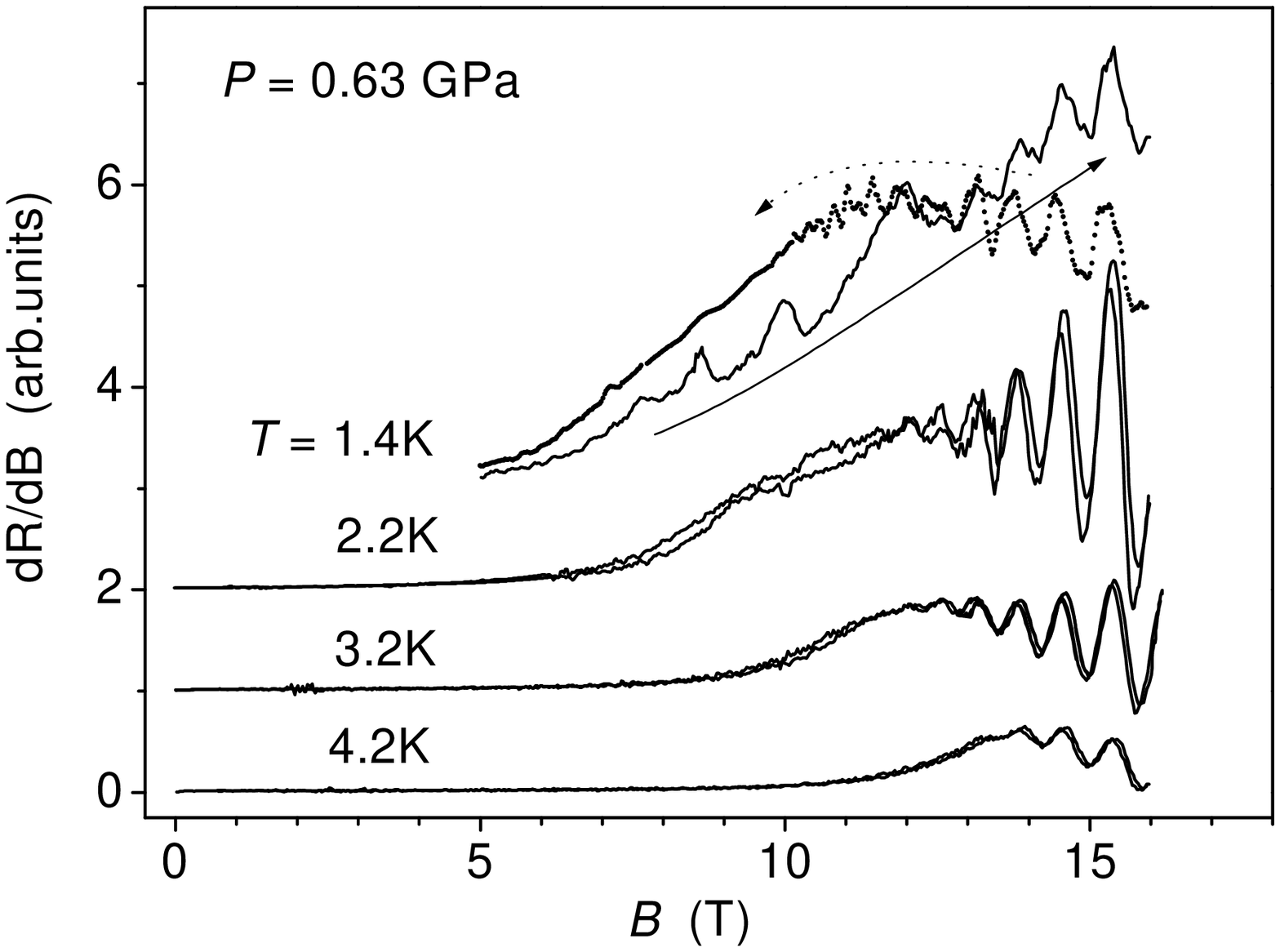,width=6.5in}}
\vspace{0.5in}

\caption{}
\label{fig4}
\end{figure}

\newpage
\begin{figure}
\centerline{\psfig{figure=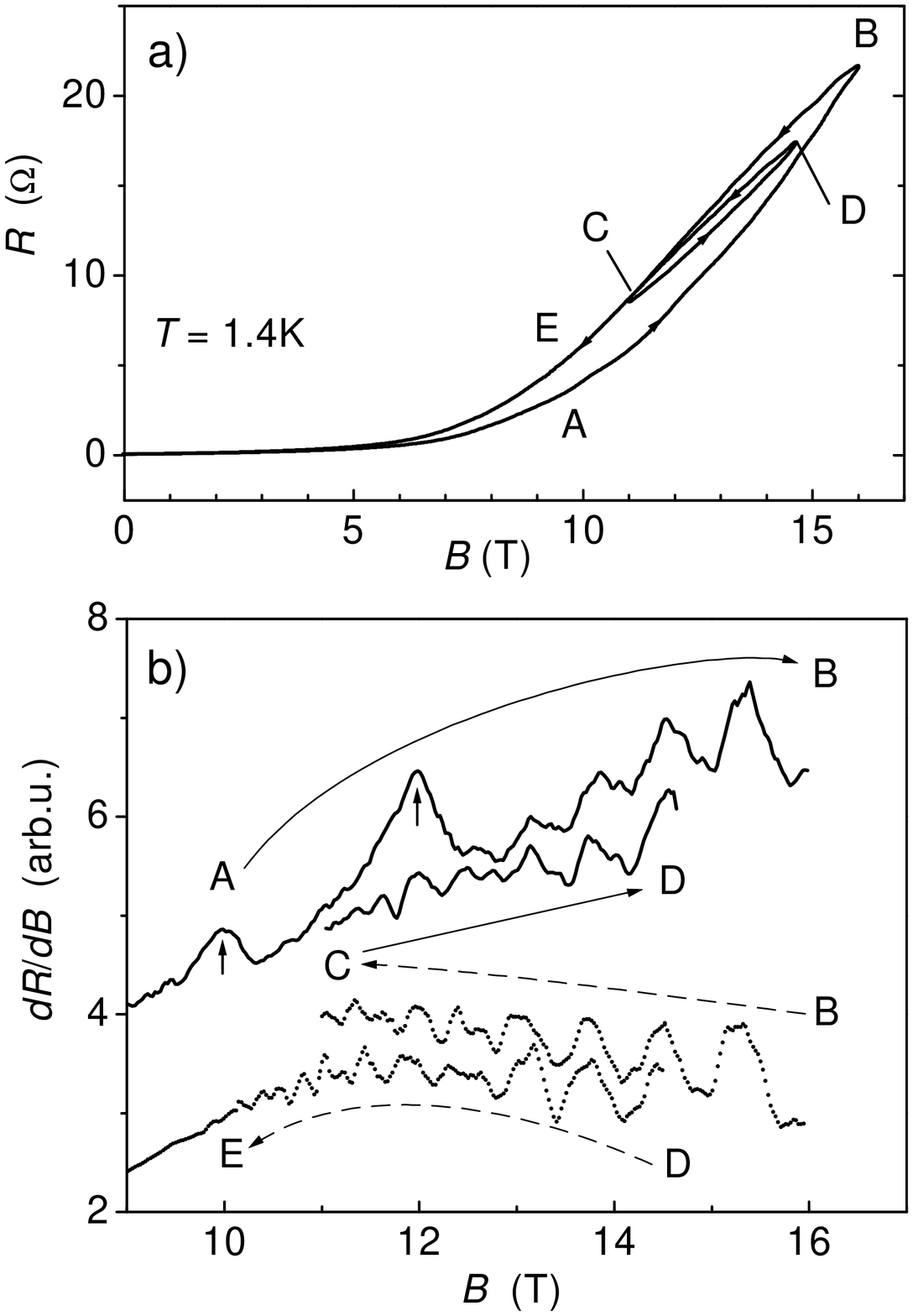,width=5.8in}}
\vspace{0.3in}
\caption{}
\label{fig5}
\end{figure}

\newpage
\begin{figure}
\centerline{\psfig{figure=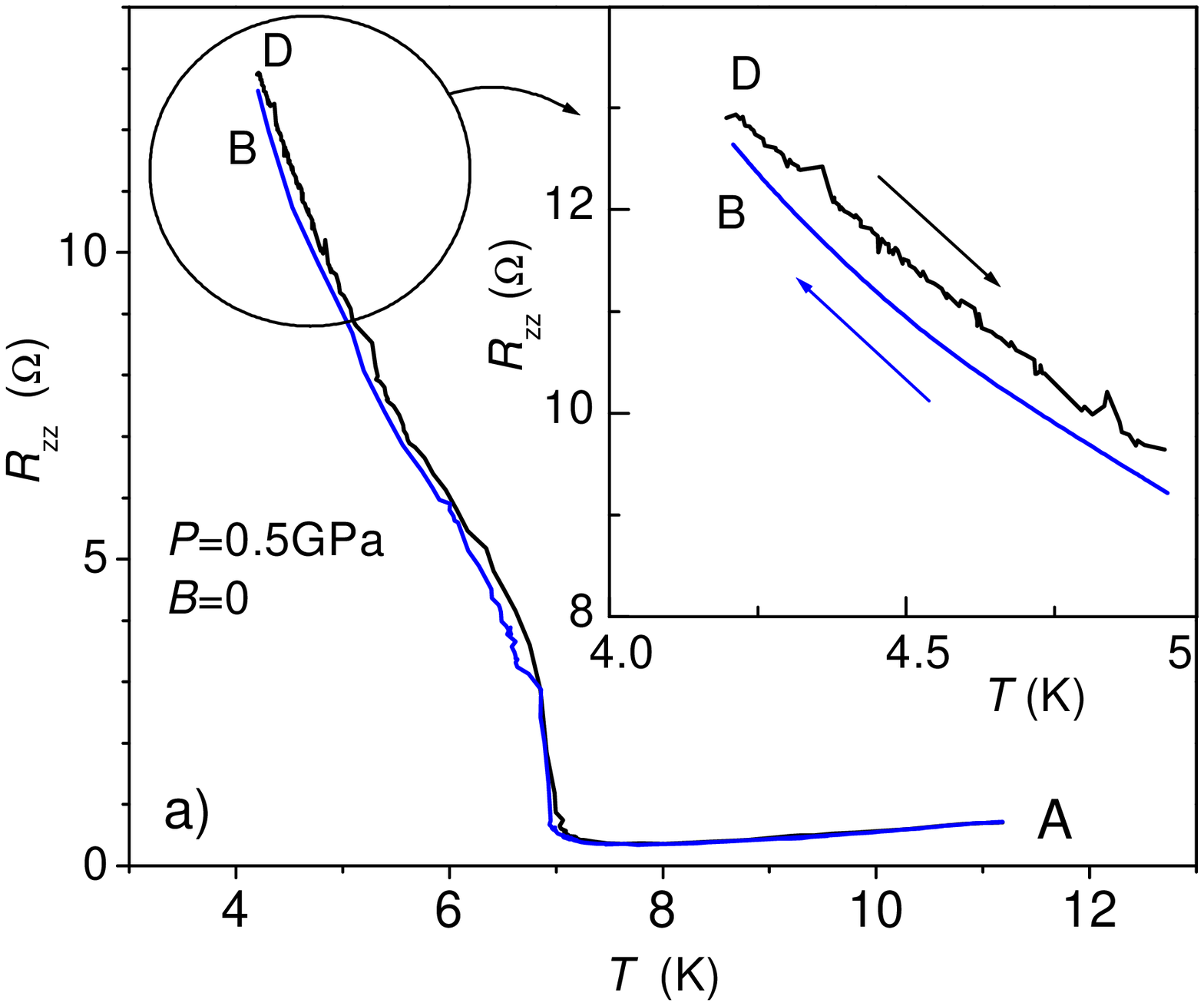,width=5.0in}} \vspace{0.5in}
\centerline{\psfig{figure=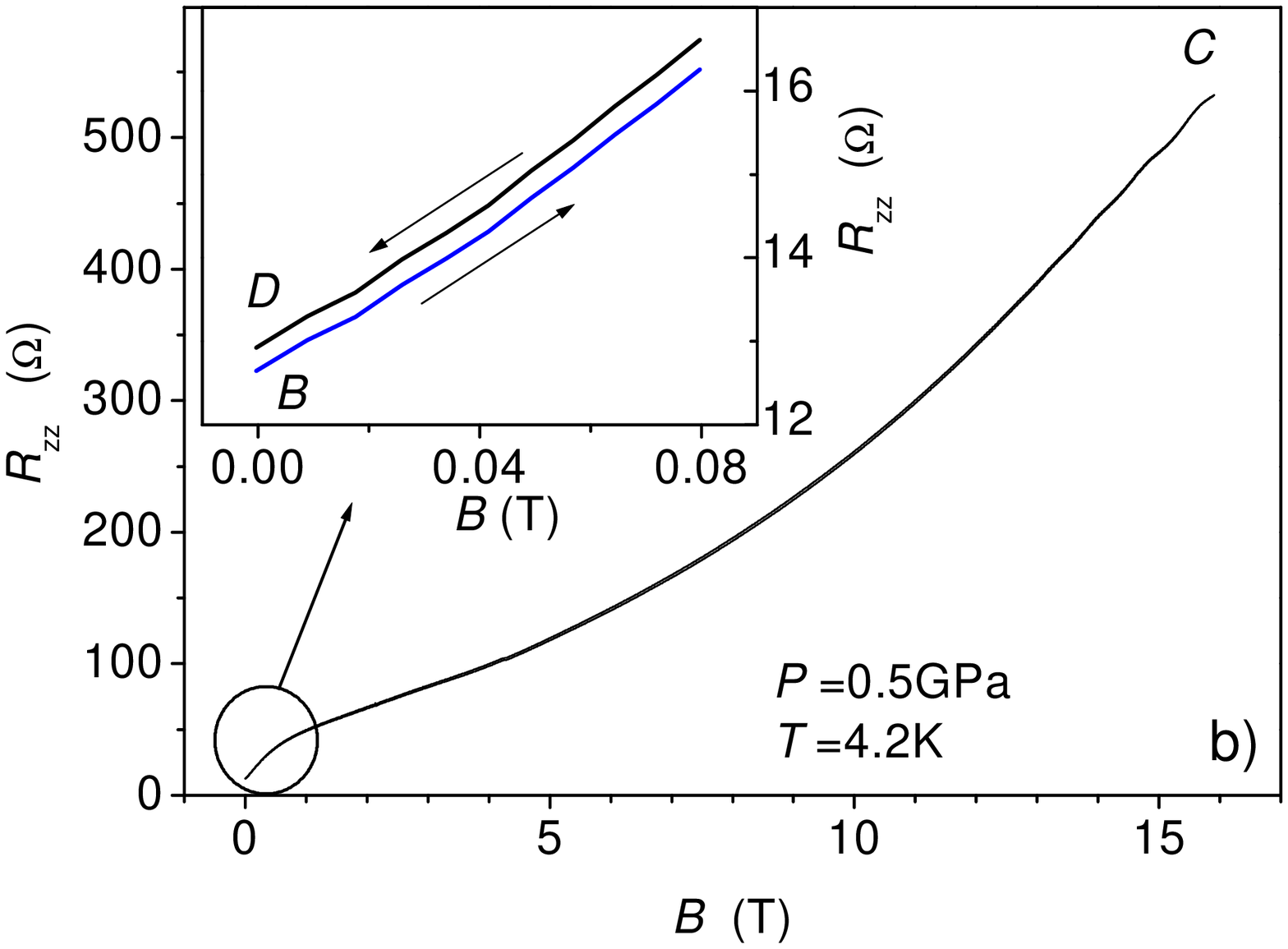,width=5.2in}} \vspace{0.2in}
\caption{} \label{fig6}
\end{figure}


\begin{references}
\bibitem{note_PM} For simplicity, throughout the paper
by ``paramagnetic'' metallic we  mean  a state at $P>P_0(T)$;
strictly speaking, this state is ``paramagnetic'' only at zero
and low magnetic field. In the quantizing magnetic field,
field-induced spin-density waves (FISDW)  arise in this region of
the phase space \protect\cite{cooper89,hannahs89}.
\bibitem{note_RO} The so called ``rapid oscillations'' (RO) are
observed in (TMTSF)$_2$PF$_6$ only in the SDW (i.e. AF) phase
in strong fields; their origin is still discussed.
For the purposes of the current paper
it is important only that for the one-dimensional Fermi surface,
RO oscillations (i) have much shorter period
than the quantized nesting oscillations related to the Fermi surface non-nested area
(and are not reduced to Shubnikov-de Haas oscillations), and (ii) have a
non-monotonic temperature dependence of the amplitude. Due to these features,
RO in (TMTSF)$_2$PF$_6$ are a good indicator of the AF phase \protect\cite{uji97} .
\end{references}
\end{document}